\documentclass[submission,copyright,creativecommons]{eptcs}
\providecommand{\event}{ACL2 workshop 2018} % Name of the event you are submitting to

\usepackage{amssymb}
\usepackage{graphicx}
\usepackage{tikz}
\usetikzlibrary{matrix}

\usepackage{breakurl}             % Not needed if you use pdflatex only.
\usepackage{underscore}           % Only needed if you use pdflatex.

\title{Formalising Filesystems in the ACL2 Theorem Prover:\\ an
  Application to FAT32}
\author{Mihir Parang Mehta
\institute{Department of Computer Science\\
University of Texas at Austin\\
Austin, TX, USA}
\email{mihir@cs.utexas.edu}}
\def\titlerunning{FAT32 Formalisation}
\def\authorrunning{M. P. Mehta}
\begin{document}
\maketitle

\begin{abstract}
In this work, we present an
approach towards constructing executable specifications of existing
filesystems and verifying their functional properties in a theorem
proving environment. We detail an application of this approach to the
FAT32 filesystem.

We also detail the methodology used to build up this type of
executable specification through a series of models which
incrementally add features of the target filesystem. This methodology
has the benefit of allowing the verification effort to start from
simple models which encapsulate features common to many filesystems
and which are thus suitable for reuse.
\end{abstract}

\section{Introduction and overview}

Filesystems are ubiquitous in computing, providing application
programs a means to store data persistently, address data by names
instead of numeric indices, and communicate with other programs.
Thus, the vast majority of application programs
directly or indirectly rely upon filesystems, which makes filesystem
verification critically important. Here, we present a
formalisation effort in ACL2 for the FAT32 filesystem, and a proof of
the read-over-write properties for FAT32 system calls. By starting
with a high-level abstract model and adding more filesystem features
in successive models, we are able to manage the complexity of this
proof, which has not, to our knowledge, been previously
attempted. Thus, this paper contributes an implementation of several
Unix-like system calls for FAT32, formally verified against an
abstract specification and tested for binary compatibility by means of
co-simulation.

In the rest of this paper, we describe these filesystem
models and the properties proved, with examples; we proceed to a
high-level explanation of these proofs and the co-simulation
infrastructure; and further we offer some insights about the low-level
issues encountered while working out the proofs.
%% I'm not sure I want to keep the statistics. Certainly ACL2 folks
%% don't want to see them.
%% We end with some statistics pertaining to the magnitude of the
%% proof effort and the running time of the proofs.

\section{Related work}

Filesystem verification research has largely followed a pattern of
synthesising a new filesystem based on a specification chosen for its
ease in proving properties of interest, rather than faithfulness to an
existing filesystem. Our work, in contrast, follows the FAT32
specification closely. In spirit, our work is closer to previous work
which uses interactive theorem provers and explores deep functional
properties than to efforts which use non-interactive theorem provers
such as Z3 to produce fully automated proofs of simpler properties.

\subsection{Interactive theorem provers}
An early effort in the filesystem verification domain was by Bevier
and Cohen~\cite{bevier1996executable}, who specified the Synergy
filesystem and created an executable model of the same in ACL2
\cite{kaufmann2000}, down to the level of processes
and file descriptors. They certified their model
to preserve well-formedness of their data structures through their
various file operations; however, they did not attempt to prove
read-over-write properties or crash consistency. Later,
Klein et al. with the SeL4 project~\cite{klein2009sel4} used
Isabelle/HOL~\cite{nipkow2002isabelle} to verify a microkernel;
while their microkernel design excluded file operations in order to keep
their trusted computing base small, it did serve as a precursor to their
more recent COGENT project~\cite{amani2016cogent}. Here the authors
built a verifying compiler to translate a filesystem specification in
their domain-specific language to C-language code, accompanied by
a proof of the correctness of this translation. Elsewhere, the SibylFS
project~\cite{ridge2015sibylfs}, again using Isabelle/HOL, provided
an executable specification for filesystems at a level of abstraction
that could function across multiple operating systems including OSX
and Unix. The Coq prover \cite{bertot2013interactive} has been used to
aid the development of FSCQ \cite{DBLP:conf/usenix/ChenZCCKZ16}, a
state-of-the art filesystem built to have high performance and
formally verified crash consistency properties.

\subsection{Non-interactive theorem provers}
Non-interactive theorem provers such as Z3 \cite{de2008z3}
have also been used to analyse filesystem models. Hyperkernel
\cite{Nelson:2017:HPV:3132747.3132748} is a recent effort which
simplifies the xv6~\cite{cox6xv6} microkernel until the point where Z3
can verify its various properties with its SMT solving
techniques. However, towards this end, all system calls in Hyperkernel
are replaced with analogs which can terminate in constant time; while
this approach is theoretically sound, it increases the chances of
discrepancies between the model and
the implementation which may diminish the utility of the proofs or
even render them moot. A stronger effort in the same domain is
Yggdrasil~\cite{sigurbjarnarson2016push}, which focusses on verifying
filesystems with the use of Z3. While the authors have made substantial
progress in terms of the number of filesystem calls they support and
the crash consistency guarantees they provide, they are subject to
the limits of SMT solving which prevent them from modelling filesystem
features such as extents, which are essential to FAT32 and many other
filesystems.

\section{Program architecture}

Modern filesystems, in response to the evolution of end users' needs
over time, have developed a substantial amount of complexity in order
to serve file operations in a fast and reliable manner. In order to
address this complexity in a principled way, we choose to build our
filesystem models incrementally, adding filesystem features in each
new model and proving equivalence with earlier, simpler models.

We have two concrete models for the FAT32 filesystem - \texttt{M2},
which is a faithful representation of a FAT32 disk image in the form
of a stobj~\cite{boyer2002single}, and \texttt{M1}, which represents the
state of the filesystem as a directory tree. This allows us to
address the practical details of updating a disk image in \texttt{M2},
which benefits from the efficient array operations ACL2 provides
for stobjs, and abstract them away in \texttt{M1} for easier reasoning
without the syntactic constraints imposed on stobj arrays.

\begin{figure}
  \centering
  \caption{Refinement/reuse relationships between abstract models}
  \label{refinement-figure}
  \begin{tikzpicture}[sibling distance=15em,
      every node/.style = {shape=rectangle, rounded corners,
        draw, align=center,
        top color=white, bottom color=blue!20}]
    \node {L1 - tree}
    child { node {L2 - length}
      child { node {L3 - unbounded disk}}
      child { node {L4 - bounded disk with garbage collection}
        child { node {L5 - permissions}}
        child { node {L6 - file allocation table}}}};
  \end{tikzpicture}
\end{figure}
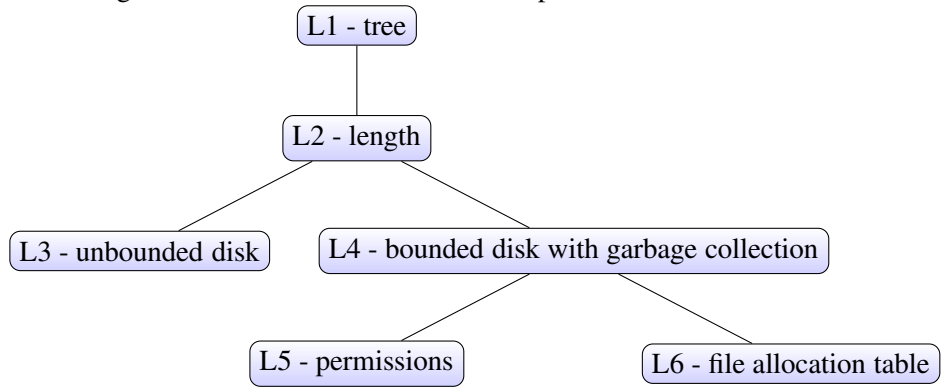

These concrete filesystem models are based upon abstract models
\texttt{L1} through \texttt{L6}. Incremental construction allows us to
reuse read-over-write proofs for simpler models in more complex
models. In the case of models \texttt{L4} and
\texttt{L6}, we are able to show a refinement relationship
without stuttering \cite{abadi1991existence}; however, for the other
models we are able to reuse proofs without proving a formal refinement
relation. These reuse relationships are summarised in figure
\ref{refinement-figure}. Much of the code and proof infrastructure is
also shared between the abstract models and the concrete models by
design. Details of the filesystem features introduced in the abstract
models can be seen in table \ref{abstract-model-description-table}.

\begin{table}[]
  \centering
  \caption{Abstract models and their features}
  \label{abstract-model-description-table}
  \begin{tabular}{|l|p{120mm}|}
    \hline
    \texttt{L1} & The filesystem is represented as a tree, with leaf
    nodes for regular files and non-leaf nodes for
    directories. The contents of regular files are represented as
    strings stored in the nodes of the tree; the storage available for
    these is unbounded. \\ \hline
    \texttt{L2} & A single element of metadata, \textit{length}, is
    stored within each regular file.  \\ \hline
    \texttt{L3} & The contents of regular files are divided into
    blocks of fixed size. These blocks are stored in an external
    ``disk'' data structure; the storage for these blocks remains
    unbounded. \\ \hline
    \texttt{L4} & The storage available for blocks is now bounded. An
    allocation vector data structure is introduced to help allocate
    and garbage-collect blocks. \\ \hline
    \texttt{L5} & Additional metadata for file ownership and access
    permissions is stored within each regular file. \\ \hline
    \texttt{L6} & The allocation vector is replaced by a file
    allocation table, matching the official FAT specification. \\ \hline
  \end{tabular}
\end{table}

A design choice that arises in this work pertains to the level of
abstraction: how operating-system specific do we want to be in our
model? Choosing, for instance, to make our filesystem operations
conform to the \texttt{file\_operations} interface~\cite{lkmpgchap4}
provided by the
Linux kernel for its filesystem modules would make our work less
general, but avert us from having to recreate some of the filesystem
infrastructure provided by the kernel. We, however, choose to
implement a subset of the POSIX filesystem application
programming interface, in order to enable us to easily compare the
results of running filesystem operations on \texttt{M2} and the Linux
kernel's implementation of FAT32, which in turn allows us to test our
implementation's correctness through co-simulation in addition to
theorem proving. As a trade-off for this choice, we are required to
implement process tables and file tables, which we do through a
straightforward approach similar to that used in
Synergy~\cite{bevier1996executable}.

At the present moment, we have implemented the POSIX system calls
\texttt{lstat}~\cite{kerrisklstat}, \texttt{open}~\cite{kerriskopen},
\texttt{pread}~\cite{kerriskpread},
\texttt{pwrite}~\cite{kerriskpwrite},
\texttt{close}~\cite{kerriskclose},
\texttt{mkdir}~\cite{kerriskmkdir} and
\texttt{mknod}~\cite{kerriskmknod}. Wherever
\texttt{errno}~\cite{kerriskerrno} is to be
set by a system call, we abide by the Linux convention.

\section{The FAT32 filesystem}
\label{sec:fat32fs}

FAT32 was initially developed at Microsoft in order to address the
capacity constraints of the DOS filesystem. Microsoft's specification
for FAT32~\cite{microsoft2000}, which we follow closely in our work,
details the layout of data and metadata in a valid FAT32 disk image.

In FAT32 all files, including regular files and directory files, are
divided into \textit{clusters} (sometimes called \textit{extents}) of
a fixed size. The size of a cluster, along with other volume-level
metadata, is stored in a \textit{reserved area} at the beginning of
the volume, and the clusters themselves are stored in a \textit{data
  region} at the end of the volume. Between these two on-disk data
structures, the volume has one or more redundant copies of the
\textit{file allocation table}.

The cluster size must be an integer multiple of the sector
size, which in turn must be at least 512 bytes; these and other
constraints on the fields of the reserved area are detailed in the
FAT32 specification.

Directory
files are for the most part treated the same way as regular files by
the filesystem, but they differ in a metadata attribute, which
indicates that the contents of directory files should be treated as
sequences of directory entries. Each such directory entry is
32 bytes and contains metadata including name, size, first
cluster index, and access times for the corresponding file.

The file allocation table is a table with one entry for each cluster
in the data region; it contains a number of linked lists
(\textit{clusterchains}). It
maps each cluster index used by a file to either the next cluster
index for that file or a special end-of-clusterchain value. \footnote{
There is a range of end-of-clusterchain values in the
specification, not just one. We support all values in the range.} This
allows the contents of a file to be reconstructed by
reading just the first cluster index from the corresponding directory
entry, reconstructing the clusterchain using the table, and then
looking up the contents of these clusters in the data
region. Unused clusters are mapped to 0 in the table; this fact is
used for counting and allocating free clusters.

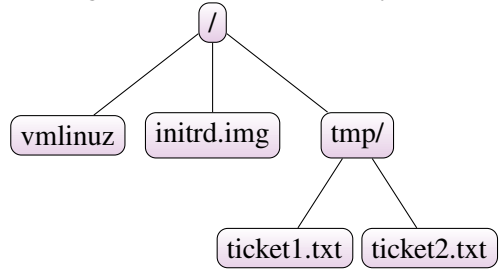
\begin{figure}
  \centering
  \caption{A FAT32 directory tree}
  \label{fat32-example}
  \begin{tikzpicture}[sibling distance=5em,
      every node/.style = {shape=rectangle, rounded corners,
        draw, align=center,
        top color=white, bottom color=violet!20}]
    \node {/}
    child { node {vmlinuz}}
    child { node {initrd.img}}
    child { node {tmp/}
      child { node {ticket1.txt}}
      child { node {ticket2.txt}}};
  \end{tikzpicture}

  \bigskip

  \begin{tabular}{|c|c|}
    \hline
    FAT index & FAT entry \\ \hline
    0 (reserved) &  \\ \hline
    1 (reserved) &  \\ \hline
    2 & \textit{EOC} \\ \hline
    3 & 4 \\ \hline
    4 & \textit{EOC} \\ \hline
    5 & \textit{EOC} \\ \hline
    6 & \textit{EOC} \\ \hline
    7 & \textit{EOC} \\ \hline
    8 & \textit{EOC} \\ \hline
    9 & 0 \\ \hline
    \vdots & \vdots
  \end{tabular}\\

  \bigskip

  \begin{tabular}{|c|c|}
    \hline
       & Directory entry in / \\ \hline
    0  & ``vmlinuz'', 3 \\ \hline
    32 & ``initrd.img'', 5 \\ \hline
    64 & ``tmp'', 6 \\ \hline
    \vdots & \vdots
  \end{tabular}

  \bigskip

  \begin{tabular}{|c|c|}
    \hline
       & Directory entry in /tmp/ \\ \hline
    0  & ``ticket1'', 7 \\ \hline
    32 & ``ticket2'', 8 \\ \hline
    \vdots & \vdots
  \end{tabular}
\end{figure}

We illustrate the file allocation table and data layout for a small
example directory tree in figure~\ref{fat32-example}. Here,
\texttt{/tmp} is a subdirectory of the root directory
(\texttt{/}). For the purposes of illustration, all regular files and
directories in this example are assumed to span one cluster except for
\texttt{/vmlinuz} which spans two clusters ($3$ and $4$), and
\textit{EOC} refers to an ``end of clusterchain'' value. Also, as
shown in the figure, the specification requires the first two entries
in the file allocation table ($0$ and $1$) to be considered reserved,
and thus unavailable for clusterchains.

\section{Proof methodology}

Broadly, we characterise the filesystem
operations we offer as either \textit{write} operations, which do
modify the filesystem, or \textit{read} operations, which do not. In
each model, we have been able to prove \textit{read-over-write}
properties which show that write operations have
their effects made available immediately for reads at the same
location, and also that they do not affect reads at other locations.

The first read-over-write theorem states that immediately following a
write of some text at some location, a read of the same length at the
same location yields the same text. The second read-over-write
theorem states that after a write of some text at some location, a
read at any other location returns exactly what it would have returned
before the write. As an example, listings for the \texttt{L1} versions
of these theorems follow. Note, the hypothesis \texttt{(stringp
  (l1-stat path fs))} stipulates that a regular file should exist at
\texttt{path}, where \texttt{l1-stat} is a function for traversing a
directory tree to look up a regular file or a directory at a given
path (represented as a list of symbols, one for each subdirectory/file
name).

\bigskip

\noindent
\begin{verbatim}
(defthm l1-read-after-write-1
  (implies (and (l1-fs-p fs)
                (stringp text)
                (symbol-listp path)
                (natp start)
                (equal n (length text))
                (stringp (l1-stat path fs)))
           (equal (l1-rdchs path (l1-wrchs path fs start text) start n) text)))

(defthm l1-read-after-write-2
  (implies (and (l1-fs-p fs)
                (stringp text2)
                (symbol-listp path1)
                (symbol-listp path2)
                (not (equal path1 path2))
                (natp start1)
                (natp start2)
                (natp n1)
                (stringp (l1-stat path1 fs)))
           (equal (l1-rdchs path1 (l1-wrchs path2 fs start2 text2) start1 n1)
                  (l1-rdchs path1 fs start1 n1))))
\end{verbatim}

By composing these properties, we can reason about executions
involving multiple reads and writes, as illustrated in the following
throwaway proof.

\medskip

\noindent
\begin{verbatim}
(thm
 (implies (and (l1-fs-p fs)
               (stringp text1)
               (stringp text2)
               (symbol-listp path1)
               (symbol-listp path2)
               (not (equal path1 path2))
               (natp start1)
               (natp start2)
               (stringp (l1-stat path1 fs))
               (equal n1 (length text1)))
          (equal (l1-rdchs path1
                           (l1-wrchs path2 (l1-wrchs path1 fs start1 text1)
                                     start2 text2)
                           start1 n1)
                 text1)))
\end{verbatim}

\begin{figure}
  \centering
  \caption{l2-wrchs-correctness-1}
  \label{l2-wrchs-correctness-1}
  \begin{tikzpicture}
    \matrix (m) [matrix of math nodes,row sep=3em,column sep=4em,minimum width=2em]
            {
              l2 & l2' \\
              l1 & l1' \\};
            \path[-stealth]
            (m-1-1) edge node [left] {l2-to-l1-fs} (m-2-1)
            edge node [below] {write} (m-1-2)
            (m-2-1.east|-m-2-2) edge node [below] {write} (m-2-2)
            (m-1-2) edge node [left] {l2-to-l1-fs} (m-2-2);
  \end{tikzpicture}
\end{figure}
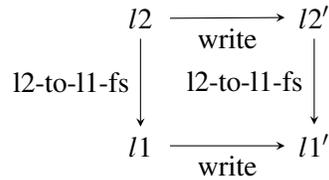

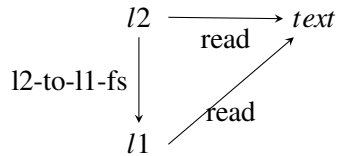
\begin{figure}
  \centering
  \caption{l2-rdchs-correctness-1}
  \label{l2-rdchs-correctness-1}
  \begin{tikzpicture}
    \matrix (m) [matrix of math nodes,row sep=3em,column sep=4em,minimum width=2em]
            {
              l2 & text \\
              l1 \\};
            \path[-stealth]
            (m-1-1) edge node [left] {l2-to-l1-fs} (m-2-1)
            edge node [below] {read} (m-1-2)
            (m-2-1.east|-m-2-2) edge node [below] {read} (m-1-2);
  \end{tikzpicture}
\end{figure}

\begin{figure}
  \centering
  \caption{l2-read-after-write-1}
  \label{l2-read-after-write-1}
  \begin{tikzpicture}
    \matrix (m) [matrix of math nodes,row sep=3em,column sep=4em,minimum width=2em]
            {
              l2 & l2' & text \\
              l1 & l1' \\};
            \path[-stealth]
            (m-1-1) edge node [left] {l2-to-l1-fs} (m-2-1)
            edge node [below] {write} (m-1-2)
            (m-2-1.east|-m-2-2) edge node [below] {write} (m-2-2)
            (m-1-2) edge node [left] {l2-to-l1-fs} (m-2-2)
            edge node [below] {read} (m-1-3)
            (m-2-2.east) edge node [below] {read} (m-1-3);
  \end{tikzpicture}
\end{figure}
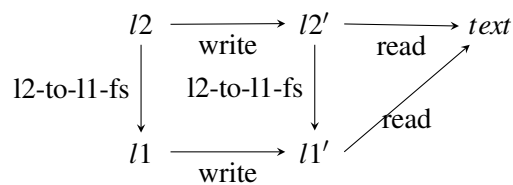

In \texttt{L1}, our simplest model, the read-over-write properties
are proven from scratch. In each subsequent model, the read-over-write
properties are proven as corollaries of equivalence proofs which
establish the correctness of read and write operations in the
respective model with respect to a previous model. A representation of
such an equivalence proof can be seen in figures
\ref{l2-wrchs-correctness-1}, \ref{l2-rdchs-correctness-1} and
\ref{l2-read-after-write-1}, which respectively show the equivalence
proof for \texttt{l2-wrchs}, the equivalence proof for
\texttt{l2-rdchs} and the composition of these to obtain the first
read-over-write theorem for model \texttt{L2}.

\section{Proof and implementation details}

We have come to rely on certain principles for the development of each
new model and its corresponding proofs. We summarise these below.

\subsection{Invariants}
\label{subsec:invariants}

As the abstract models grow more complex, with the addition of more
auxiliary data the ``sanity'' criteria for filesystem instances become more
complex. For instance, in \texttt{L4}, the predicate \texttt{l4-fs-p}
is defined to be the same as \texttt{l3-fs-p}, which recursively
defines the shape of a valid directory tree. However, we choose to
require two more properties for a ``sane'' filesystem.

\begin{enumerate}
\item Each disk index assigned to a regular file should be
  marked as \textit{used} in the allocation vector - this is essential
  to prevent filesystem errors.
\item Each disk index assigned to a regular file should be distinct
  from all other disk indices assigned to files - this does not hold
  true, for example, in filesystems with hardlinks. FAT32 lacks
  hardlinks, and we can use this fact to make our proofs easier.
\end{enumerate}

These properties are invariants to be maintained across
write operations; while not all of them are strictly necessary for a
filesystem instance to be valid, they do simplify the verification of
read-over-write properties by helping us ensure that write operations
do not create an aliasing situation in which a regular file's
contents can be modified through a write to a different regular file.

These properties, in the form of the predicates
\texttt{indices-marked-listp} and \texttt{no-duplicatesp}, are
packaged together into the \texttt{l4-stricter-fs-p} predicate, for
which a listing follows. Here, the allocation vector \texttt{alv}
satisfies the type hypothesis \texttt{(boolean-listp alv)}.

\medskip

\noindent
\begin{verbatim}
(defun l4-stricter-fs-p (fs alv)
  (declare (xargs :guard t))
  (and (l4-fs-p fs)
       (boolean-listp alv)
       (let ((all-indices (l4-list-all-indices fs)))
            (and (no-duplicatesp all-indices)
                 (indices-marked-p all-indices alv)))))
\end{verbatim}

Similarly, for proof purposes we find it useful to package together
certain invariants for the stobj \texttt{fat32-in-memory}, which we
maintain while manipulating the stobj through input/output operations
and file operations, in the predicate
\texttt{compliant-fat32-in-memoryp} for which a listing follows. The
constant \texttt{*ms-first-data-cluster*}, for instance, is $2$, and
the last clause of the conjunction helps us maintain an invariant
about all cluster indices including the root cluster being greater
than or equal to $2$, which is stipulated in the FAT32 specification
(section \ref{sec:fat32fs}) and therefore necessary while reasoning
about operations in the file allocation table.

\medskip

\noindent
\begin{verbatim}
(defund compliant-fat32-in-memoryp (fat32-in-memory)
  (declare (xargs :stobjs fat32-in-memory :guard t))
  (and (fat32-in-memoryp fat32-in-memory)
       (>= (bpb_bytspersec fat32-in-memory) *ms-min-bytes-per-sector*)
       (>= (bpb_secperclus fat32-in-memory) 1)
       (>= (count-of-clusters fat32-in-memory)
           *ms-fat32-min-count-of-clusters*)
       (>= (bpb_rootclus fat32-in-memory) *ms-first-data-cluster*)))
\end{verbatim}

\texttt{M2}, however, differs from previous models in that the
predicate \texttt{compliant-fat32-in-memoryp} cannot be assumed to
hold before the filesystem is initialised by reading a disk
image. This also means that it is not straightforward to use an
abstract stobj~\cite{goel2013abstract} for modelling the filesystem
state, since the putative invariant
\texttt{(compliant-fat32-in-memoryp fat32-in-memory)} is not always
satisfied.

\subsection{Reuse}

As noted earlier, in our abstract models, using a refinement
methodology allows us to derive our read-over-write properties
with little additional effort; more precisely, we are able to prove
read-over-write properties simply with \texttt{:use} hints after
having done the work of proving refinement through induction.

At a lower level, we are also able to benefit from refinement
relationships between components of our different models. For example,
such a relationship exists between the allocation vector used in
\texttt{L4} and the file allocation table used in \texttt{L6}. More
precisely, by taking a file allocation table and mapping each non-zero
entry to \texttt{true} and each zero entry to \texttt{false}, we
obtain a corresponding allocation vector with exactly the same amount
of available space. This is a refinement mapping which makes it a lot
easier to prove that \texttt{L4} itself is an abstraction of
\texttt{L6}. This, in turn, means that the effort spent on proving the
invariants described above for \texttt{L4} need not be replicated for
\texttt{L6}.

\subsection{The FTY discipline}

In the model \texttt{M1} in particular, we use the FTY discpline and
its associated library \cite{15-swords-fty} to simplify our
definitions for regular files,
directory files, and other data types. This allows us to simplify as
well as speed up our reasoning by eliminating many type hypotheses,
and in particular allows us to prove read-over-write properties for
\texttt{M1} with a significantly smaller number of helper lemmas
compared to our abstract models in which FTY is not used.

\subsection{Allocation and garbage collection}

In the development of our abstract models, \texttt{L4} is the first to
be bounded in terms of disk size; accordingly, data structures are
needed for allocating free blocks on disk, including those which were
previously occupied by other files. For simplicity, we choose to model
the allocation vector used by the CP/M filesystem~\cite{moriacpm},
which is a boolean
array equal in size to the disk recording whether disk blocks are free
or in use. Using this, we are able to prove that write operations
succeed if and only if there is a sufficient number of free blocks
available on the disk, and when we later model FAT32's file allocation
table in \texttt{L6}, we are able to extend this result by proving a
refinement relationship between these two allocation data structures
as described earlier.

While it is not straightforward to reuse this result for the file
allocation table in \texttt{M2}, we are still able to benefit from
reusing the \texttt{L6} algorithms for disk block allocation and
reconstruction of file contents for a given file. As a result, our
\texttt{L6} proofs for the correctness of these algorithms are also
available for reuse in \texttt{M2}.

\section{Stobjs and co-simulation}

Previous work on executable specifications \cite{goel2014simulation}
has shown the importance of testing these on real examples, in order
to validate that the behaviour shown matches that of the system being
specified. In our case, this means we must validate our filesystem by
testing it in execution against a canonical implementation of FAT32;
in this case, we choose the implementation which ships with Linux
kernel 3.10.

For each of our tests, we need to produce FAT32 disk images, which we
do with the aid of the program
\texttt{mkfs.fat}~\cite{hudsonmkfs}. Further, we make use of this
program's verbose mode (enabled through the \texttt{-v} command-line
switch) for a simple yet fundamental co-simulation test. In the verbose
mode, \texttt{mkfs.fat} emits an English-language summary of the
fields of the
newly created disk image; we use \texttt{diff}~\cite{eggertdiff} to
compare this summary against the output of an ACL2 program, based on
our model, which reads the image and pretty-prints the FAT32 fields in the
same format. This validates our code for constructing an in-memory
representation of the disk image and also serves as a regression test
during the process of modifying the model to support proofs and
filesystem calls.

Further, we co-simulate \texttt{cp}~\cite{granlundcp}, a simple
program for copying files, by
reproducing its functionality in an ACL2 program. This allows us to
validate our code for reading and writing regular files and
directories which span multiple clusters.

The quick execution of these co-simulation tests relies upon the use
of stobjs. Since the file allocation table and data region
(section~\ref{sec:fat32fs}) are large arrays which must be initialised
whenever a new disk image is read, ACL2's efficient array operations
turn out to be very helpful; this is also the case when we write back
a disk image after performing a sequence of file operations. We have
also developed a number of special-purpose macros for facilitating the
generation of lemmas needed for reasoning about reads and writes to
the fields of a stobj in the filesystem context; we expect these to
continue to be useful as we work with abstract and concrete stobjs for
modelling new filesystems.

%% This stuff is not yet implemented exactly as described.

%% Our first co-simulation test is for \texttt{cat}~\cite{granlundcat}, a
%% simple program which reads its input and copies it to its output. Its
%% functionality is reproduced in an ACL2 program which uses our
%% implementations of the system calls
%% \texttt{lstat}~\cite{kerrisklstat},
%% \texttt{open}~\cite{kerriskopen}, \texttt{pread}~\cite{kerriskpread},
%% \texttt{pwrite}~\cite{kerriskpwrite}, and
%% \texttt{close}~\cite{kerriskclose}. This allows us to validate our
%% code for reading and writing regular files and directories which span
%% multiple clusters.

%% This stuff is not yet implemented at all.

%% We further test the \texttt{dd}~\cite{rubindd} program, which provides
%% similar functionality to \texttt{cat} but with more options to
%% customise the data transfer, for instance, by allowing the data
%% transfer to be split into blocks of a specified size.

\section{Conclusion}

This work formalises a FAT32-like filesystem and proves
read-over-write properties through refinement of a series of
models. Further, it proves the correctness of FAT32's allocation and
garbage collection mechanisms, and provides artefacts to be used in a
subsequent filesystem models.

\section{Future work}

The FAT32 model is still a work in progress; the set of system calls
is not yet complete and the translation functions between disk images,
\texttt{M2} instances and \texttt{M1} instances are not yet
verified. However, many of the techniques for these proofs have
already been demonstrated in our abstract models. Once we have these,
we intend to use them as a basis for
reasoning about sequences of filesystem operations in a program, in a
manner akin to proving properties of code on microprocessor
models. This is a motivation for the pursuit of binary compatibility
in our work.

While FAT32 is interesting of and by itself, it lacks features such as
crash consistency, which most modern filesystems provide by means of
journalling. We hope to reuse some artefacts of formalising FAT32 in
order to verify a filesystem with journalling, such as
ext4~\cite{mathur2007new}.

We also hope to model the behaviour of filesystems in a
multiprogramming environment, where concurrent filesystem calls must
be able to occur without corruption or loss of data. Although we have
not yet used abstract stobjs for the reasons described
above in subsection~\ref{subsec:invariants}, we plan to explore their
use in facilitating our stobj reasoning.
%% Also, we hope to support "code proofs", by providing a basis for
%% reasoning about filesystem operations in filesystem-specific utilities
%% such as \texttt{fsck}, as well as other application programs. This is
%% a large part of the motivation for pursuing binary compatibility.

\subsubsection*{Acknowledgments} This material is based upon work
supported by the National Science Foundation SaTC program under
contract number CNS-1525472. Thanks are also owed to Warren A. Hunt
Jr. and Matt Kaufmann for their guidance, and to the anonymous
reviewers for their suggestions.

\bibliographystyle{eptcs}
\bibliography{references}
\end{document}